\input{aipcheck}

\documentclass[
    ,final            
  ]
  {aipproc}

\layoutstyle{8x11double}

\begin{document}

\title{QCD, monopoles on the lattice and gauge invariance.}

\classification{12.38.Aw, 14.80.Hv, 11.15.Ha, 11.15.Kc}
\keywords      {QCD, Lattice, Monopoles, Color confinement}

\author{C. Bonati}{
  address={Dipartimento di Fisica and INFN, Pisa, Italy}}

\author{M. D'Elia}{address={Dipartimento di Fisica and INFN, Genova, Italy}}
\author{A. Di Giacomo {\footnote{presentd the talk}}
} {address={Dipartimento di Fisica and INFN, Pisa, Italy} }

\begin{abstract}
  The number and the location of the monopoles observed on the lattice in QCD configurations
  happens to depend strongly on the choice of the gauge used to expose them, in contrast to the physical expectation that monopoles be gauge invariant objects. It is proved by use of the non abelian Bianchi identities (NABI) that monopoles are indeed gauge invariant, but the method used to detect them
  depends, in a controllable way, on the choice of the abelian projection. Numerical checks are presented. 
 \end{abstract}

\maketitle


\section{Introduction}
Monopoles play an important role in non abelian gauge theories. They can condense in the vacuum and produce dual superconductivity, which is a good candidate to be the mechanism of color confinement\cite{'tHP}\cite{m}\cite{'tH2}. The prototype monopole configuration is the soliton solution of Ref.'s \cite{'tH,Pol} in the Higgs-broken phase of an $SU(2)$ gauge theory coupled to a scalar field in the adjoint representation (Georgi-Glashow model). The property which characterizes it as a monopole is a non trivial homotopy $\pi_2$: it realizes a non trivial mapping of the sphere $S_2$ at spatial infinity onto $SU(2)/U(1)$. This property is invariant under continuous gauge transformations.

 A monopole in non-abelian gauge theory is an intrinsically abelian object: the magnetic monopole term 
in a multipole expansion obeys abelian equations and identifies a $U(1)$ subgroup of the gauge group, modulo an arbitrary global rotation $\times$ an arbitrary gauge transformation $U(\vec r)$ which tends to the identity as $r\to\infty$\cite{coleman}. In the soliton solution of Ref.\cite{'tH,Pol} this $U(1)$ are the rotations around the color direction of the vacuum expectation of the Higgs field. The corresponding abelian field tensor is known as 't Hooft tensor \cite{'tH2}. In $QCD$ there is no Higgs field and apparently no privileged residual U(1) symmetry. It was proposed in Ref.\cite{'tH2} that
 any operator $O$ in the adjoint representation can act as an effective Higgs field, the physics being for some reason independent of that choice.
Any choice of $O$ identifies an "effective" unitary representation or Abelian Projection. 

The recipe to detect monopoles in $U(1)$ lattice configurations\cite{dgt} relies on the observation that any excess over $2\pi$ of the magnetic flux through a plaquette is a Dirac string crossing it. If the net number of Dirac strings crossing the border of an elementary cube is non zero, there is a monopole inside the cube. In $U(1)$ gauge theory the magnetic flux is gauge invariant, and hence this procedure is gauge invariant and unambiguous. In non-abelian theories one first fixes a gauge (abelian projection), and then plays the same game as in $U(1)$ on the residual $U(1)$ subgroup. The result is in this case strongly dependent on the choice of the gauge, so that the existence or non-existence of a monopole in a site is a gauge dependent feature, and this is physically unacceptable. Either monopoles can be created or destroyed  by a gauge transformation, and therefore are unphysical, or the statement
of Ref.\cite{'tH2} that all abelian projections are physically equivalent is not correct.

We will solve this problem by use of the non-abelian Bianchi identities ($NABI$).
  
  \section{the non-abelian Bianchi identities}
 The $U(1)$ (abelian) Bianchi identities  $\partial_{\mu}F^*_{\mu \nu}=0$, or
$ \vec \nabla \vec B = 0 $ , $ \vec \nabla \wedge \vec E + \partial_t\vec B=0 $
  imply zero magnetic current.  A violation has the form  
  $\partial_{\mu}F^*_{\mu \nu} = j_{\nu}$.
  The magnetic current $j_{\nu}$ is conserved $\partial_{\nu} j_{\nu}=0$, because of the antisymmetry of $F^*_{\mu \nu}$.
  
The non-abelian Bianchi identities ($NABI$) read
  \begin{equation}
  D_{\mu} G^*_{\mu \nu} = J_{\nu} \label{NABI}
  \end{equation}
  a gauge covariant equation. It implies\cite{bdlp}
   \begin{equation}
   D_{\nu} J_{\nu} =0
   \end{equation}
   The four components of the magnetic current $J_{\mu}$ commute with each other\cite{cm}.

To extract the physical (gauge-invariant) information contained in the  $NABI$ Eq.(1), we can diagonalize them by a gauge transformation and project on a complete set of diagonal matrices,
 $\phi_0^a$ $(a= 1,..r)$, $r$  rank of the gauge group. 

A convenient choice for   $\phi_0^a$ are the fundamental weights  : there is one of them for each simple root of the Lie algebra $\vec \alpha^a$.

     $\left[\phi_0^a, H_i\right] =0$ , $\left[\phi_0^a,E_{\pm \vec \alpha}\right] =\pm (\vec c^a \vec \alpha)E_{\pm \vec \alpha}$, $(\vec c^a \vec \alpha^b)=\delta_{ab}$.
     
     If we call $\phi^a_{I}$ the adjoint matrix which is equal to $\phi^a_0$ in the representation in which $J_{\mu}$ is diagonal the projection gives
      \begin{equation}
       Tr(\phi_{I}^aD_{\mu}G^*_{\mu \nu})=Tr(\phi^a_{I}J_{\nu}) \equiv j_{\nu}^a(x,I)  \label{PMA}
       \end{equation}
   We can also project on the adjoint matrix   ${\phi^a_{V}}$ which is equal to $V(x)\phi_{0}^aV^{\dagger}(x)$ , [$V(x)$ any gauge transformation], in the gauge where $J_{\mu}$ is diagonal. $\phi^a_V$ coincides with   ${\phi^a_I}$in the special case $V=I$.
    \begin{equation}
Tr(\phi_{V}^aD_{\mu}G^*_{\mu \nu})=Tr(\phi_{V}^aJ_{\nu}) \equiv  j_{\nu}^a (x,V) \label{PMB}
\end{equation}
Let  ${F^a}_{\mu \nu}(x,V)$ be the 't Hooft tensor , i.e. the abelian field strength $\partial_{\mu} A^a_{\nu} - \partial_{\nu} A^a_{\mu}$ in the gauge (abelian projection) in which  $\phi^a_{V} $ is diagonal. 

In Ref.\cite{bdlp} we proved the following 

{\bf THEOREM   For a generic compact gauge group and for any V(x) Eq.(\ref{PMB}) is equivalent to
 \begin{equation}
 \partial_{\mu} {F^a}^*_{\mu \nu}(x,V) = j_{\nu}^a (x,V) \label{ME}
 \end{equation} }
 \vspace{0.4cm}
 
 The abelian Bianchi identities in a generic abelian projection are a consequence of the $NABI$.
\section{The soliton monopole revisited.}
We now check our theorem Eq.(\ref{ME}) on the soliton configuration  of Ref.\cite{'tH,Pol}. The lagrangean is

$L = -\frac{1}{4}\vec G_{\mu \nu}\vec G_{\mu \nu} + (D_{\mu} \phi)^{\dagger}(D_{\mu} \phi) -V(\phi^2)$

The hedgehog gauge is defined  as that in which the $vev$ of the Higgs field  $ \vec \phi(\vec r) = H(r) \hat r$  , $H(r)_{r \to \infty}\to v$. In this gauge the solution is

\hspace{1cm}$\vec A_0 =0$  \hspace{1.5cm}        $ A^a_i= -\epsilon_{iak} \frac{r^k}{gr^2} [ 1 - K(gvr) ] $

$[ 1 - K(x)]_{x\to 0}\propto x^2 $ ,\hspace{1.5cm}      $K(x)_{x\to \infty} \approx \exp(-x)$

This gauge is nothing but the Landau Gauge. Indeed 
 $\partial_{\mu} A_{\mu} = \partial_{i} A_{i} =0$ \hspace{1cm}.
 
 The 't Hooft tensor is \hspace{0.5cm}    $F_{ij} = \partial_{i} A^3_{j}- \partial_{j} A^3_{i} $\hspace{0.5cm}and the abelian magnetic field $b_i =\frac{1}{2}\epsilon_{ijk} F_{jk}$. Direct computation gives
   \begin{equation}
 \vec b \approx_{r \to\infty} \frac{2\hat r}{gr^2}cos(\theta)
 \end{equation}  
 \begin{equation}
 Q_m =r^2\int d\Omega \hat r \cdot \vec b =\frac{2}{g} \int d\Omega cos(\theta) =0  \!  \!
 \end{equation}
 No magnetic charge !
 
In the unitary gauge: \hspace {0.4cm}  $\phi_{i}\sigma_i = v \sigma_{3}$. The solution is static and hence  $J_{i}=0$ $(i=1,..3)$. For the charge density
\begin{equation}
J_{0} = D_{i}B_{i} = \frac{2\pi}{g}\delta^3(\vec r)\sigma_3
\end{equation}
By direct calculation \cite{bdlp} it is found that in this gauge the Maximal Abelian gauge condition satisfied

\hspace{2.1cm} $\partial_{\mu}A_{\mu}^{\pm} \pm ig \left[A_{\mu}^3, A_{\mu}^{\pm} \right] =0$ 

In $SU(2)$ there is only one fundamental weight  $\phi_0={\sigma_3\over 2}$ (Rank of $SU(2)$  =1) .

Our theorem Eq.\eqref{ME} gives
\begin{equation}
Tr(\phi_0\vec D \cdot \vec B )= \vec \nabla \cdot \vec b = Tr({\phi_0 J_0})=\frac{2\pi}{g} \delta^3(\vec r)
\end{equation}
or
\begin{equation}
Q_m= \frac{1}{g} 
\end{equation}
 We can compute the magnetic charge in  gauges interpolating between maximal abelian and Landau
 acting on the maximal abelian with
 \begin{equation}
  U_{{\bf a}}(\theta, \phi) = \exp(i\phi\frac{\sigma_{3}}{2}) \exp(i\theta {\bf a} \frac{\sigma_{2}}{2}) \exp(-i\phi\frac{\sigma_{3}}{2})  \label{gt}
  \end{equation}

 For ${\bf a}=0$ $U_{0}(\theta, \phi)=1$, and one stays in the maximal abelian, for ${\bf a}=1$  $U_{1}(\theta, \phi)$ transforms to Landau gauge. 
    
   Direct computation gives \cite{bdd}
  \begin{equation}
  \frac{Q_{m}({\bf a})}{Q_{m}(0)} =\frac{1+ \cos({\bf a} \pi)}{2}  \le 1 \label{qratio}
  \end{equation}
  The magnetic charge measured by the flux at infinity is maximum in the maximal abelian and decreases to zero approaching the Landau gauge.
  
  By use of the analysis of Ref.\cite{coleman} one can show that the above analysis holds not only for
  the soliton monopole, but for a generic configuration, which can be viewed as a point-like magnetic charge producing the monopole field at large distances plus a background of zero total magnetic charge 
  responsible for higher multipoles\cite{bdlp}.
  
  What we have discussed for the gauge group $SU(2)$ can be extended to any gauge theory, with and without coupling to quark fields \cite{bdlp} \cite{dlp}. In the following we shall therefore omit the index $^a$  in the currents and in the fields since in $SU(2)$ there is only one fundamental weight.
  \section{Confinement}
  An order parameter for dual superconductivity is the $vev$ of a gauge invariant operator $O(x)$ carrying magnetic charge:  if $\langle O \rangle \neq 0$ the magnetic $U(1)$ is Higgs broken and there is superconductivity (confinement), whilst in the deconfined phase $\langle O \rangle = 0$\cite{digz}.
  
  The magnetic charge density operator in the maximal-abelian gauge corresponds to $V(x)=I$ in Eq.\eqref{ME}, i.e. to the current defined by Eq. \eqref{PMA}, $j_{0}(x,I) = Tr[\phi_{I}J_{0}(x)]$; the magnetic charge is $Q_{I}=\int d^3xj_{0}(x,I)$. Suppose  that $\langle O(x)\rangle\neq 0$ and
$[Q ,O(x)]= mO(x)$ $m\neq 0$ i.e. that the residual $U(1)$ in the maximal abelian gauge is Higgs-broken.  As an element of the Lie algebra $V(x)\phi_{0}V^{\dagger}(x)=C(V,x)\phi_{0} + \sum_{\vec \alpha}C_{\vec \alpha}(V,x)E_{ \vec \alpha}$ with $\vec \alpha$ the roots of the algebra. Computing $j_{0}(V,x) = Tr[V(x)\phi_{0}V^{\dagger}(x)J_{0}(x)]$ in the representation where $J_{0}$ is diagonal one gets $j_{0}(x,V)= C(V,X)j_{0}(x,I)$ and
 \begin{equation} 
  [Q_{V},O(x)] = C(V,x) O(x)
  \end{equation}
  Since $C(V,x)$ is generically non zero $O(x)$ is magnetically charged also with respect to the
  $U(1)$ selected by the Higgs field $V(x)\phi_{0}V^{\dagger}(x)$, and also the corresponding $U(1)$ is 
  Higgs broken. 
  Dual superconductivity is abelian projection invariant.
   
  \section{Numerical check on the lattice}
  To check the idea that monopoles are gauge invariant objects, and that it is the detection method 
  of Ref.\cite{dgt} which is abelian projection dependent, we produce by standard methods  an ensemble of $SU(2)$ configurations in the maximal abelian gauge and we look for monopoles \cite{bdd}.
  We identify the elementary cubes containing a monopole, and for the sake of simplicity we only consider the cubes with one single Dirac string crossing the border: in principle there could be many of them, corresponding e.g. to strings going through the cube, but we expect them to become irrelevant in the continuum limit. We then assume that the monopole is at the centre of the cube and that the direction of the string is perpendicular to the face crossed, which we assume as 3-rd axis. We then perform the gauge transformation Eq.\eqref{gt} depending on the parameter $\alpha$ and  we measure the magnetic charge by the method of Ref.\cite{dgt}
  The result is shown in Fig.(1), for different values of the  coupling $\beta\equiv \frac{4}{g^2} $ and compared to the prediction Eq.\eqref{qratio}.
  \begin{figure}[ht]
\vspace{1mm}
\scalebox{0.28}{\rotatebox{0}{\includegraphics{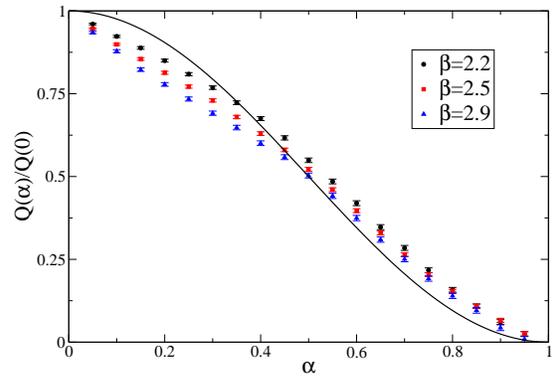}}}
\caption{Results and theoretical prediction (black line) for the ratio defined in Eq.\eqref{qratio}.}
\label{ris_fig}
\end{figure}

The agreement is very good if we consider the systematic errors due to discretization \cite{bdd}, which are, among others, responsible for the slight $\beta$ dependence observed.  

This is a clear demonstration that monopole existence is a gauge invariant fact: observing it in different abelian projections by the recipe of Ref.\cite{dgt} gives a probability of detection which is always less than $1$, and is equal to $1$  in the maximal abelian gauge. It is not correct to speak of monopoles 
in different abelian projections as if they were different objects. In particular it is not true that no monopoles exist in the Landau gauge: the correct statement is that the monopoles are gauge independent, but they escape detection with the recipe of Ref.\cite{dgt} in the Landau gauge.

In conclusion, by use of the non abelian Bianchi identities, we have shown that the existence of a monopole in a gauge field configuration is a gauge invariant concept. In a magnetically charged configuration  a privileged direction in color space exists, that of the magnetic monopole term
in the multipole expansion of the field, which is selected by the maximal abelian gauge.
 
 Confinement is a projection independent property.







\bibliographystyle{aipproc}   

\bibliography{sample}

\IfFileExists{\jobname.bbl}{}
 {\typeout{}
  \typeout{******************************************}
  \typeout{** Please run "bibtex \jobname" to optain}
  \typeout{** the bibliography and then re-run LaTeX}
  \typeout{** twice to fix the references!}
  \typeout{******************************************}
  \typeout{}
 }


\end{document}